\documentclass[twocolumn]{aastex631}
\usepackage[utf8]{inputenc}

\usepackage{amsmath}

\usepackage{xcolor}
\usepackage{color}
\definecolor{orange}{rgb}{0.90,0.60,0}
\definecolor{skyblue}{rgb}{0.35,0.70,0.90}
\definecolor{green}{rgb}{0,0.60,0.50}
\definecolor{yellow}{rgb}{0.95,0.90,0.25}
\definecolor{blue}{rgb}{0,0.45,0.70}
\definecolor{vermilion}{rgb}{0.80,0.40,0}
\definecolor{lilac}{rgb}{0.80,0.60,0.70}

\newcommand{\dif}{\mathrm{d}}
\newcommand{\e}{\mathrm{e}}

\newcommand{\Myr}{\,\mathrm{Myr}}
\newcommand{\Gyr}{\,\mathrm{Gyr}}
\newcommand{\AU}{\,\mathrm{AU}}
\newcommand{\pc}{\,\mathrm{pc}}
\newcommand{\ME}{M_{\oplus}}
\newcommand{\RE}{R_{\oplus}}

\newcommand{\MSol}{M_{\odot}}
\newcommand{\RSol}{R_{\odot}}

\newcommand{\yuavg}{\langle y_{u} \rangle}
\newcommand{\yzavg}{\langle y_{z} \rangle}
\newcommand{\perGyrstar}{\Gyr^{-1} \, {\rm star}^{-1}}

\defcitealias{Spina+2021}{S21}
\defcitealias{Behmard+2023b}{B23}
\defcitealias{Liu+2024}{L24}

\begin{document}

\title{Metal pollution in Sun-like stars from destruction of ultra-short-period planets}

\author[0000-0003-3987-3776]{Christopher E.\ O'Connor}
\affiliation{Center for Interdisciplinary Exploration and Research in Astrophysics (CIERA), Northwestern University, 1800 Sherman Ave, Evanston, IL 60201, USA}
\affiliation{Department of Astronomy and Cornell Center for Astrophysics and Planetary Science, Cornell University, Ithaca, NY 14853, USA}
\correspondingauthor{Chris O'Connor}
\email{christopher.oconnor@northwestern.edu}

\author[0000-0002-1934-6250]{Dong Lai}
\affiliation{Department of Astronomy and Cornell Center for Astrophysics and Planetary Science, Cornell University, Ithaca, NY 14853, USA}
\affiliation{Tsung-Dao Lee Institute, Shanghai Jiao Tong University, Shanghai 201210, China}

\shorttitle{Tidal destruction of USPs}
\shortauthors{O'Connor \& Lai}

\submitjournal{AAS Journals}

\begin{abstract}
    Chemical evidence indicates that an appreciable fraction of Sun-like stars 
    have engulfed rocky planets during their main-sequence lifetimes. 
    We investigate whether the tidal evolution and destruction of ultra-short-period planets (USPs) can explain this phenomenon. 
    We develop a simple parameterized model for the formation and engulfment of USPs in a population of MS stars. 
    With this model, it is possible to reproduce both the observed occurrence rate of USPs and the frequency of planet-engulfing Sun-like stars 
    for a reasonable range of USP formation rates and tidal decay lifetimes. 
    Our results support a theory of USP formation through gradual inward migration over many Gyr 
    and suggest that engulfment occurs $\sim 0.1$--$1 \Gyr$ after formation. 
    This lifetime is set by tidal dissipation in the USP itself instead of the host star, 
    due to the perturbing influence of external companions. 
    If USP engulfment is the main source of pollution among Sun-like stars, 
    we predict a correlation between pollution and compact multi-planet systems; 
    some $5$--$10\%$ of polluted stars should have a transiting planet of mass $\gtrsim 5 \ME$ and period $\sim 4$--$12$ days.  
    We also predict an anti-correlation between pollution and USP occurrence. 
\end{abstract}

\keywords{Exoplanet dynamics (490), Exoplanet tides (497), Extrasolar rocky planets (511), Star-planet interactions (2177)}

\section{Introduction} \label{s:Intro} 

Short-period exoplanets are potentially vulnerable to tidal disruption and engulfment by their host stars. 
This may happen as a result of stellar radius expansion, tidal orbital decay, or a combination of the two. 
The orbital decay of short-period gas giants, or hot Jupiters (HJs), is inferred 
from studies of the rotation rates and age distribution of their host stars 
\citep[e.g.][]{Penev+2018, Qureshi+2018, HS2019, TejadaArevalo+2021, MiyazakiMasuda2023}; 
it has also been detected directly in one system \citep{Yee+2020}. 

Separate evidence indicates that small, rocky exoplanets are frequently ingested by Sun-like main sequence (MS) stars as well. 
Detailed studies of chemical composition among co-natal stellar pairs -- stars with a common origin -- 
reveal unexpectedly large differential abundances among refractory elements \citep[e.g.,][]{Ramirez+2015, Oh+2018, TucciMaia2019}. 
In many systems, these differences are best explained if the metal-enriched star in a pair engulfed a rocky planet of $\sim 1$--$10 \ME$ well into its MS lifetime; 
statistical analyses find signs of this in $\sim 3$--$30 \%$ of co-natal Sun-like stars \citep{Spina+2021, Behmard+2023b, Liu+2024}. 
We refer to this as ``pollution,'' in reference to a similar phenomenon among white dwarfs \citep[e.g.,][]{Koester2014}.

If rocky planet engulfment is so widespread, how does it come about? 
Many forms of violent dynamical evolution are possible in planetary systems, each potentially able to inject a planet into the star. 
However, based on dynamical injection probabilities and observational planetary occurrence rates, 
one predicts at most $\sim 2 \%$ of single FGK dwarfs being polluted as a result of all violent mechanisms combined (Appendix \ref{app:violent}). 
As this falls below the inferred fraction of single stars polluted by planets ($[8 \pm 3] \%$ from \citealt{Liu+2024}), 
violent dynamics seems unlikely to be the dominant route to engulfment.

Another possibility is that the engulfed planets were originally ultra-short-period planets (USPs), 
small ($\leq 2 \RE$) worlds with orbital periods $P \leq 1$ day. 
The properties of USPs naturally suggest a connection with the observed pollution: 
They are mainly rocky worlds with typical masses of $\sim 0.5$--$8 \ME$ \citep{SanchisOjeda+2014, Uzsoy+2021}. 
This matches the refractory composition and total amount of accreted material inferred in polluted stars \citep{Spina+2021, Liu+2024}. 
Moreover, USPs are distributed in orbital period from 1 day down to the critical period for tidal disruption,
\begin{equation}
    P_{\rm RLO} = 4.1 \, {\rm h} \left( \frac{\eta}{2} \right)^{3/2} \left( \frac{\bar{\rho}}{5 \, {\rm g \, cm^{-3}}} \right)^{-1/2},
\end{equation}
where $\bar{\rho}$ is the planet's mean density and $\eta$ is a factor depending on its internal structure \citep{Eggleton1983, Rappaport+2013}. 
This implies that tidal evolution is important in their histories \citep{PDW2019, PL2019, MS2020}. 
A few USPs are either actively disintegrating \citep{Rappaport+2012_disintegratingUSP, Rappaport+2014_disintegratingUSP, SanchisOjeda+2015_disintegratingUSP} 
or possibly tidally distorted \citep{Dai+2024}.
An apparent drawback of the hypothesized connection with pollution is that USPs are rare, occurring around $0.5\%$ of FGK dwarfs \citep{SanchisOjeda+2014, Uzsoy+2021}. 
However, this does not preclude the possibility that polluted stars once had USPs and subsequently engulfed them.

In this Letter, we pursue the hypothesis that planetary pollution among Sun-like stars 
is linked with the tidal evolution and destruction of USPs. 
In Section \ref{s:USPFormEvol}, we summarize current theories of USP formation and dynamical evolution. 
In Section \ref{s:Model}, we develop a simple analytical model 
tracking the formation and destruction of USPs as a population. 
With it, we calculate the average USP formation rate and the typical lifetime before engulfment 
required to reproduce the observed occurrence rates of USPs and engulfed planets. 
We also demonstrate the compatibility of our results with USP formation models and tidal evolution. 
In Section \ref{s:Discuss}, we discuss observational implications of our results and a few caveats. 
We conclude in Section \ref{s:Conclusion}.

\section{USP formation} \label{s:USPFormEvol}

Many properties of USPs can be explained readily if they formed as the innermost members 
of compact super-Earth systems like those discovered by {\it Kepler}  
and migrated inward through planet--planet interactions and tidal friction \citep{Schlaufman2010}.
Three main scenarios have been proposed: 
\begin{enumerate}
    \item {\it High-eccentricity (high-e) migration}, where a proto-USP reaches a small periastron distance with a high eccentricity such that tidal circularization occurs rapidly. 
    One avenue for this is ``secular chaos'' in systems of three or more widely spaced planets with moderate eccentricities and inclinations \citep{PDW2019}. 

    \item {\it Low-eccentricity (low-e) migration}, where secular interactions in compact systems of three or more planets keep a proto-USP's eccentricity at a moderate value ($\sim 0.1$), 
    thereby enabling gradual tidal migration \citep{PL2019}. 

    \item {\it Obliquity-driven migration}, where a proto-USP has its obliquity excited by a companion through capture into a secular spin--orbit resonance \citep{MS2020}. 
    Tidal migration is rapid but ends abruptly when the planet escapes the resonance \citep[see also][]{SuLai2022}.
\end{enumerate}
In this work, we consider low-e migration when making quantitative estimates. 
We shall discuss how the other two scenarios lead to different predictions. 

\section{Population model} \label{s:Model}

\subsection{Definitions and key equations} \label{s:Model:Formulation}

Consider an ensemble of Sun-like stars, of which a fraction $y_{p0}$ is born 
with planetary systems whose architectures are conducive to USP production. 
Let the fraction of stars of age $t$ that currently host a USP be $y_{u}(t)$, 
and let the fraction that have previously engulfed a USP be $y_{z}(t)$. 
We compute the time evolution of these quantities, making the following assumptions: 
\begin{enumerate}
    \item Each planet-hosting star can form at most one USP during its MS lifetime. 
    This is consistent with the observed architectures of compact multi-planet systems, 
    where only the innermost planet has a chance to undergo significant orbital decay during the MS.
    
    \item The formation of USPs is described by a ``source function'' $\Sigma(t)$ 
    giving the number of planets reaching $P = 1$\,day per planet-hosting star per unit time. 
    The fraction of stars that form a USP between ages $t$ and $t+\dif t$ is then $y_{p}(t) \Sigma(t) \, \dif t$, where
    \begin{subequations}
    \begin{align}
        y_{p}(t) &= y_{p0} \exp\left( - \int_{0}^{t} \Sigma(t') \, \dif t' \right) \\
        &= y_{p0} - y_{u}(t) - y_{z}(t)
    \end{align}
    \end{subequations}
    is the fraction of stars at age $t$ that have not yet formed a USP.
    
    \item After reaching $P = 1$\,day, a USP undergoes tidal decay and is engulfed after a time $\tau$, 
    such that the number of stars that engulf a USP between ages $t$ and $t+\dif t$ is $(1/\tau) y_{u}(t) \, \dif t$.
    
    \item After engulfing a USP, the host star's atmosphere is polluted with refractory elements for the rest of its MS lifetime.
\end{enumerate}
Under these assumptions, $y_{u}(t)$ and $y_{z}(t)$ obey 
\begin{subequations} \label{eq:diffeqs_yu_yz}
\begin{align}
    \frac{\dif y_{u}}{\dif t} &= - \frac{y_{u}(t)}{\tau} + \left[ y_{p0} - y_{u}(t) - y_{z}(t) \right] \Sigma(t), \label{eq:diffeqs_yu_yz:yu} \\
    \frac{\dif y_{z}}{\dif t} &= \frac{y_{u}(t)}{\tau}. \label{eq:diffeqs_yu_yz:yz}
\end{align}
\end{subequations}
These equations can also be derived from a continuity equation 
describing a population of inward-migrating planets. 

\subsection{Solution with constant source function} \label{s:Model:ConstSrcFunc}

The simplest case has a constant source function, $\Sigma(t) = \Sigma_{0}$, 
describing sustained USP formation over stellar lifetimes. 
We solve equations (\ref{eq:diffeqs_yu_yz}) for $t \geq 0$ with $y_{u}(0) = y_{z}(0) = 0$, obtaining
\begin{subequations} \label{eq:yu_yz_srcconst}
\begin{align}
    y_{u}(t) &= \frac{y_{p0} \Sigma_{0} \tau}{1 - \Sigma_{0} \tau} \left( \e^{-\Sigma_{0} t} - \e^{-t/\tau} \right), \label{eq:yu_yz_srcconst:yu} \\
    y_{z}(t) &= y_{p0} \left[ 1 - \left( \frac{\e^{-\Sigma_{0} t} - \Sigma_{0} \tau \e^{-t/\tau}}{1 - \Sigma_{0} \tau} \right) \right]. \label{eq:yu_yz_srcconst:yz}
\end{align}
\end{subequations}
We consider $\Sigma_{0} \tau \ll 1$, as we expect $y_{u} \ll 1$ for all $t$. 

It is instructive to examine the limits $t \ll \tau$ and $t \gg \tau$:
\begin{equation} \label{eq:yu_limits}
    y_{u}(t) \simeq
    \left\{
    \begin{array}{ll}
        y_{p0} \Sigma_{0} t, & t \ll \tau; \\
        y_{p0} \Sigma_{0} \tau e^{-\Sigma_{0} t}, & t \gg \tau
    \end{array}
    \right.
\end{equation}
and
\begin{equation} \label{eq:yz_limits}
    y_{z}(t) \simeq
    \left\{
    \begin{array}{ll}
        y_{p0} \Sigma_{0} t^{2} / (2 \tau) & t \ll \tau; \\
        y_{p0} (1 - e^{-\Sigma_{0} t}), & t \gg \tau.
    \end{array}
    \right.
\end{equation}
In particular, for $\tau \ll t \ll 1/\Sigma_{0}$:
\begin{subequations} \label{eq:yu_yz_quasisteady}
\begin{align} 
    y_{u}(t) &\simeq y_{p0} \Sigma_{0} \tau, \\
    y_{z}(t) &\simeq y_{p0} \Sigma_{0} t.
\end{align}
\end{subequations}
In this regime, the USP population is in a quasi-steady state with occurrence rate $\simeq y_{p0} \Sigma_{0} \tau$, 
while the occurrence rate of engulfed planets grows linearly with time. 
Depending on $\Sigma_{0}$ and $\tau$, it is possible for a significant fraction of MS stars 
to engulf a USP even though USPs are rare for stars of any given age.

\begin{figure}
    \centering
    \includegraphics[width=\columnwidth]{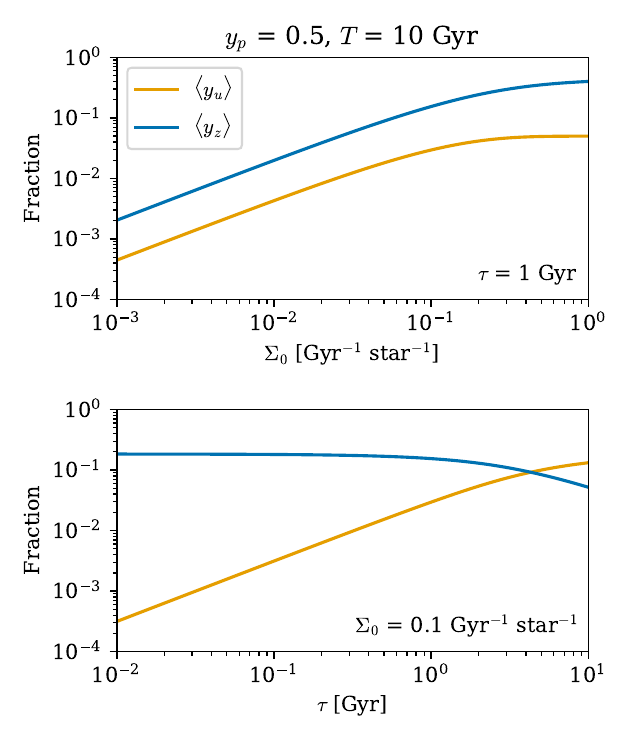}
    \caption{Variation of $\yuavg$ (orange curves) and $\yzavg$ (blue) versus $\Sigma_{0}$ and $\tau$ (lower)
    under the constant-source-function population model (equations \ref{eq:define_yuavg_yzavg}). 
    In each panel, the other model parameter is held fixed at the value indicated in the lower right corner. 
    Note in particular the simple behaviors that occur for $\Sigma_{0} \tau \ll 1$.}
    \label{fig:yuyz_Sig0Tau}
\end{figure}

Although highly reductive, this model is useful for illustrating the conditions under which 
a population of tidally decaying USP systems can reproduce pollution in Sun-like stars. 
We suppose that the ensemble is composed of $1 \MSol$ stars with ages uniformly distributed between $0$ and $10 \Gyr \equiv T$ 
-- a reasonable first approximation of the general population of G-dwarfs in the Galaxy. 
We assume that a fraction $y_{p0} = 0.5$ of stars can host USPs, the same occurrence rate as ``Kepler multis'' \citep{Fressin+2013, Petigura+2013, Zhu+2018, Mulders+2018, Hsu+2019}. 
Given $\Sigma_{0}$ and $\tau$, we calculate the observed fractions of stars with USPs or pollution 
as the time-averaged values of equations (\ref{eq:yu_yz_srcconst}):
\begin{subequations} \label{eq:define_yuavg_yzavg}
\begin{align} 
    \yuavg &\equiv \frac{1}{T} \int_{0}^{T} y_{u}(t) \, \dif t \\
    &= \frac{y_{p0} \tau}{T} \left[ 1 - \left( \frac{\e^{-\Sigma_{0} T} - \Sigma_{0} \tau \e^{-T/\tau}}{1 - \Sigma_{0} \tau} \right) \right], \label{eq:define_yuavg_yzavg:yuavg_formula} \\
    \yzavg &\equiv \frac{1}{T} \int_{0}^{T} y_{z}(t) \, \dif t \\
    &= y_{p0} \left[ 1 - \left( \frac{ 1 - \e^{-\Sigma_{0} T} - \Sigma_{0}^{2} \tau^{2} \left( 1 - \e^{-T/\tau} \right)}{\Sigma_{0} T (1 - \Sigma_{0} \tau)} \right) \right]. \label{eq:define_yuavg_yzavg:yzavg_formula}
\end{align}
\end{subequations}
Figure \ref{fig:yuyz_Sig0Tau} shows how $\yuavg$ and $\yzavg$ vary with respect to each parameter
for $\Sigma_{0} \in [10^{-3}, 1] \perGyrstar$ and $\tau \in [10^{-2},10] \Gyr$. 
While varying one parameter, we hold the other fixed at a fiducial value $\tau = 1.0 \Gyr$ or $\Sigma_{0} = 0.1 \perGyrstar$. 
We see that both $\yuavg$ and $\yzavg$ increase linearly in proportion with $\Sigma_{0}$ -- 
the higher the USP formation rate, the greater number of USPs and engulfment events -- 
before leveling off for $\Sigma_{0} \gtrsim 0.1 \perGyrstar$. 
Similarly, we find $y_{u} \propto \tau$ at fixed $\Sigma_{0}$ before again reaching a plateau for $\tau \gtrsim 10 \Gyr$. 
On the other hand, $\yzavg$ is nearly constant for $\tau \lesssim 1 \Gyr$ and decreases somewhat for large $\tau$.

\subsection{Results} \label{s:Model:ConstSrcFunc:Obs}

We now calculate the $\Sigma_{0}$ and $\tau$ corresponding 
to the observed occurrence rates of USPs and metal-polluted Sun-like stars. 
We adopt a total USP occurrence rate of $\yuavg = 0.0048$ around Sun-like stars \citep{SanchisOjeda+2014, Uzsoy+2021}, 
and we set $y_{p0} = 0.5$ and $T = 10 \Gyr$ unless otherwise specified. 
The three most recent studies of the rate of rocky planet engulfment among FGK dwarfs 
are \citet[][S21]{Spina+2021}, \citet[][B23]{Behmard+2023b}, and \citet[][L24]{Liu+2024}. 
In Table \ref{tab:observations_to_Sig0_tau}, we list their reported values (interpreted as $\yzavg$) and the corresponding $\Sigma_{0}$ and $\tau$ from equations (\ref{eq:define_yuavg_yzavg}). 
The derived $\Sigma_{0}$--$\tau$ pairs span roughly an order of magnitude in each quantity. 
We reproduce the observed $\yuavg$ and $\yzavg$ for $\Sigma_{0} \approx 0.01$--$0.2 \perGyrstar$ and $\tau \approx 0.1$--$0.8 \Gyr$. 
Using moderately different values of $y_{p0}$ has little effect on $\tau$ 
but changes $\Sigma_{0}$ roughly as $\Sigma_{0} \propto y_{p0}^{-1}$. 

\begin{figure}
    \centering
    \includegraphics[width=\linewidth]{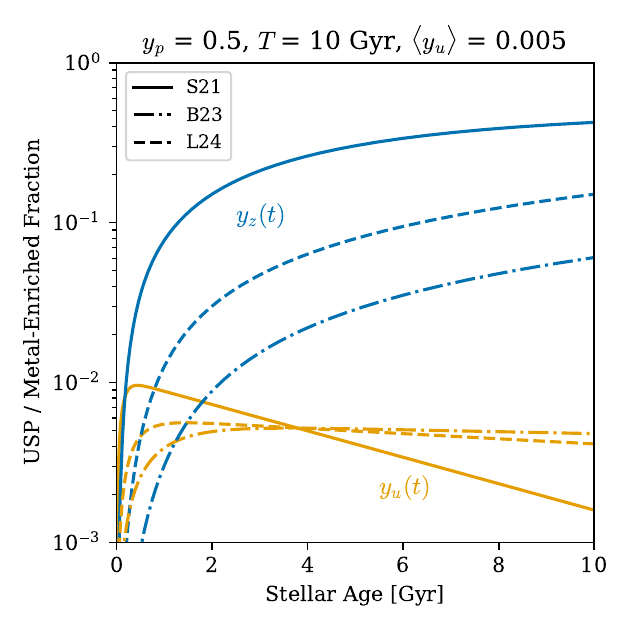}
    \caption{Time evolution of $y_{u}$ (orange curves) and $y_{z}$ under the constant-source-function model, 
    using parameters derived from \citet[][solid curves]{Spina+2021}, 
    \citet[][dot-dashed]{Behmard+2023b}, and \citet[][dashed]{Liu+2024}.}
    \label{fig:yuyz_evolution}
\end{figure}

In Figure \ref{fig:yuyz_evolution}, we graph $y_{u}(t)$ and $y_{z}(t)$ for each parameter pair in Table \ref{tab:observations_to_Sig0_tau}. 
For the S21-based model, $y_{u}$ grows rapidly at early ages, peaks around $t \sim 0.3 \Gyr$, 
then declines exponentially on a characteristic time $1/\Sigma_{0} \approx 5 \Gyr$. 
The B23- and L24-based models predict that the USP frequency grows at a moderate pace among young stars 
before leveling off at $t \sim 1$--$2 \Gyr$ and declining only slightly by $t = 10 \Gyr$. 
These trends are described well by equations (\ref{eq:yu_limits}) through (\ref{eq:yu_yz_quasisteady}). 

A useful metric for comparison between models is the average age of USP host stars,
\begin{equation}
    \langle t_{*u} \rangle = \frac{\int_{0}^{T} t \, y_{u}(t) \, \dif t}{\int_{0}^{T} y_{u}(t) \, \dif t}.
\end{equation} 
We list calculated values of $\langle t_{*u} \rangle$ for each model in Table \ref{tab:observations_to_Sig0_tau}. 
The B23- and L24-based models both predict $\langle t_{*u} \rangle \approx 5 \Gyr$, 
consistent the Galactic stellar population as a whole. 
The S21-based model predicts a moderately younger $\langle t_{*u} \rangle = 3.6 \Gyr$. 
For completeness, we also calculate the average age of metal-enriched stars, $\langle t_{*z} \rangle$, in the same way as $\langle t_{*u} \rangle$. 
We find $\langle t_{*z} \rangle \approx 6$--$7 \Gyr$ in all three cases (Table \ref{tab:observations_to_Sig0_tau}). 

We next evaluate whether the derived $\Sigma_{0}$ and $\tau$ values 
are compatible with theories of USP formation and tidal evolution. 

\begin{deluxetable*}{llcccc} \label{tab:observations_to_Sig0_tau}
\tablecaption{Frequency of rocky planet engulfment among Sun-like stars ($\yzavg$) as reported in various recent studies, 
and derived quantities as calculated using the constant-source-function population model 
with $\yuavg = 0.0048$, $y_{p0} = 0.5$, and $T = 10 \Gyr$.}
\tablewidth{0pt}
\tablehead{ \colhead{Source} & \colhead{$\yzavg$} & \colhead{$\Sigma_{0}$} & \colhead{$\tau$} & \colhead{$\langle t_{*u} \rangle$} & \colhead{$\langle t_{*z} \rangle$} \\
    & & [Gyr$^{-1}$ star$^{-1}$] & [Gyr] & [Gyr] & [Gyr]
    }
\startdata
    \citet{Behmard+2023b} & 0.029 & 0.014 & 0.78 & 5.3 & 6.9 \\
    \citet{Liu+2024} & 0.077 & 0.037 & 0.32 & 4.9 & 6.7 \\
    \citet{Spina+2021} & 0.27 & 0.19 & 0.11 & 3.6 & 6.3 
\enddata
\end{deluxetable*}

\subsection{Comparison with USP formation theory} \label{s:Model:CompareUSPForm}

USP formation theories predict that $\Sigma_{0}$ is given by the rate of orbital migration due to tidal dissipation,
starting from an orbital period of $P_{1i} > 1$ day. 
In the low-e migration scenario \citep{PL2019}, $P_{1i} \approx 1$--$3$ days 
and the USP formation rate is given by
\begin{subequations} \label{eq:Sigma0_PL2019}
\begin{align}
    \Sigma_{0} \sim \left| \frac{\dot{a}_{1}}{a_{1}} \right| &= \frac{7 e_{1}^{2}}{2 t_{a1}} \\
    &= 21 k_{2,1} \Delta t_{\rm L,1} \frac{G M_{\star}^{2} R_{1}^{5}}{m_{1} a_{1i}^{8}} e_{1}^{2}  \\
    &\approx 0.08 \perGyrstar \left( \frac{k_{2,1} \Delta t_{\rm L,1}}{100 \, {\rm s}} \right) \nonumber \\
    & \hspace{0.5cm} \times \left( \frac{R_{1}}{\RE} \right)^{5} \left( \frac{m_{1}}{\ME} \right)^{-1} \left( \frac{e_{1}}{0.01} \right)^{2} \nonumber \\
    & \hspace{0.5cm} \times \left( \frac{M_{\star}}{\MSol} \right)^{-2/3} \left( \frac{P_{1i}}{2 \, {\rm day}} \right)^{-16/3}. \label{eq:Sigma0_PL2019_numbers}
\end{align}
\end{subequations} 
The quantities $m_{1}$, $R_{1}$, $k_{2,1}$, and $\Delta t_{\rm L1}$ are the proto-USP's 
mass, radius, real-valued tidal Love number, and tidal lag time; 
$e_{1}$ is its eccentricity. 
The timescale $t_{a1}$ characterizes the overall dissipation in the planet. 

High-e or obliquity-driven migration would predict qualitatively similar results \citep{PDW2019, MS2020}, 
although the latter invokes a finite planetary obliquity rather than eccentricity to drive orbital decay. 
Comparing equation (\ref{eq:Sigma0_PL2019_numbers}) to the values of $\Sigma_{0}$ in Table \ref{tab:observations_to_Sig0_tau}, 
we see our constraint on $\Sigma_{0}$ is broadly consistent with USP formation theory, 
regardless of which specific migration channel dominates. 

\subsection{Comparison with tidal evolution theory} \label{s:Model:CompareTidalEvo}

In general, a planet's orbital decay rate includes contributions from tidal dissipation in both the planet itself and the host star. 
If the planet has a nearly circular orbit (semi-major axis $a_{1}$, eccentricity $e_{1}$), its rotation is pseudo-synchronized, and the host star rotates slowly,
then the rate of semi-major axis decay is
\begin{equation} \label{eq:adot_div_a_tides}
    \frac{\dot{a}_{1}}{a_{1}} = - \left( \frac{1}{t_{a\star}} + \frac{7 e_{1}^{2}}{2 t_{a1}} \right).
\end{equation} 
The tidal evolution timescale for stellar dissipation is:
\begin{equation}
    \frac{1}{t_{a\star}} = \frac{9}{2 Q'_{\star}} \frac{m_{1}}{M_{\star}} \left( \frac{R_{\star}}{a_{1}} \right)^{5} \left( \frac{G M_{\star}}{a_{1}^{3}} \right)^{1/2}, \label{eq:define_ta:star}
\end{equation}
where $Q'_{\star}$ is the star's reduced quality factor. 
We assume a constant $Q'_{\star} = 3 \times 10^{6}$.  
This is consistent with the conditions under which \citet{PL2019} were able to reproduce the USP period distribution, 
as well as empirical estimates of dissipation in exoplanet hosts \citep[e.g.][]{Hansen2010, Hansen2012, Penev+2018}. 

Tidal evolution is dominated by the planet's dissipation when the eccentricity exceeds a critical value:
\begin{subequations} \label{eq:define_e1cr}
\begin{align}
    e_{\rm 1cr} & \equiv \left( \frac{2}{7} \frac{t_{a1}}{t_{a\star}} \right)^{1/2} \\
    &= \left( \frac{3}{14 k_{2,1} \Delta t_{\rm L1} n_{1} Q'_{\star}} \right)^{1/2} \frac{m_{1}}{M_{\star}} \left( \frac{R_{\star}^{5}}{R_{1}^{5}} \right)^{5/2} \\
    &\approx 0.002 \left( \frac{k_{2,1} \Delta t_{\rm L1}}{100 \, {\rm s}} \frac{Q'_{\star}}{3 \times 10^{6}} \right)^{-1/2} \left( \frac{P_{1}}{1 \, {\rm day}} \right)^{1/2} \nonumber \\
    & \hspace{0.5cm} \times \left( \frac{m_{1}}{\ME} \right) \left( \frac{M_{\star}}{\MSol} \right)^{-1} \left( \frac{R_{\star}}{\RSol} \right)^{5/2} \left( \frac{R_{1}}{\RE} \right)^{-5/2}.
\end{align}
\end{subequations}
In the low-e migration scenario, a nonzero eccentricity is maintained by secular forcing in a compact multi-planet system. 
We calculate the forced eccentricity due to the innermost companion 
with mass $m_{2}$, semi-major axis $a_{2}$, and r.m.s.\ eccentricity $e_{2}$ \citep{PL2019}: 
\begin{equation} \label{eq:e1forced}
    e_{\rm 1f} = \frac{5}{4} \frac{a_{1}}{a_{2}} \frac{e_{2}}{1 + \varepsilon_{\rm tide} + \varepsilon_{\rm GR}}
\end{equation}
The quantities $\varepsilon_{\rm tide}$ and $\varepsilon_{\rm GR}$ measure the short-range forces 
created by the planet's tidal bulge and GR corrections to the stellar potential:
\begin{subequations}
\begin{align}
    \varepsilon_{\rm tide} &= \frac{10 k_{2,1} M_{\star}^{2} R_{1}^{5} a_{2}^{3}}{m_{1} m_{2} a_{1}^{8}} \\
    &\approx 0.2 \left( \frac{k_{2,1}}{0.3} \right) \left( \frac{M_{\star}}{\MSol} \right)^{2} \left( \frac{m_{1}}{\ME} \frac{m_{2}}{10 \ME} \right)^{-1} \left( \frac{R_{1}}{\RE} \right)^{5} \nonumber \\
    &\hspace{0.5cm} \times \left( \frac{a_{2}}{0.1 \AU} \right)^{3} \left( \frac{a_{1}}{0.02 \AU} \right)^{-8}, \nonumber \\
    \varepsilon_{\rm GR} &= \frac{4 G M_{\star}^{2} a_{2}^{3}}{c^{2} m_{2} a_{1}^{4}} \label{eq:define_epsGR} \\
    &\approx 8 \left( \frac{M_{\star}}{M_{\odot}} \right)^{2} \left( \frac{m_{2}}{10 \ME} \right)^{-1} \left( \frac{a_{2}}{0.1 \, {\rm AU}} \right)^{3} \left( \frac{a_{1}}{0.02 \, {\rm AU}} \right)^{-4}. \nonumber
\end{align} 
\end{subequations}

The USP's inspiral lifetime from an initial semi-major axis $a_{1i}$ may be expressed as 
\begin{align} \label{eq:tau_lifetime_estimate}
    \tau = f \left| \frac{\dot{a}_{1}}{a_{1}} \right|^{-1}_{a_{1} = a_{1i}},
\end{align}
where $f$ is a factor of order unity. 
If only stellar tides are active, then $f = 2/13 \approx 0.154$. 
Through numerical integration of equation (\ref{eq:adot_div_a_tides}), 
we find $f \approx 0.2$--$0.8$ gives a better estimate for us, 
due to the additional contributions of planetary tides and SRFs. 
We use $f = 0.5$ going forward. 

\begin{figure}
    \centering
    \includegraphics[width=0.5\textwidth]{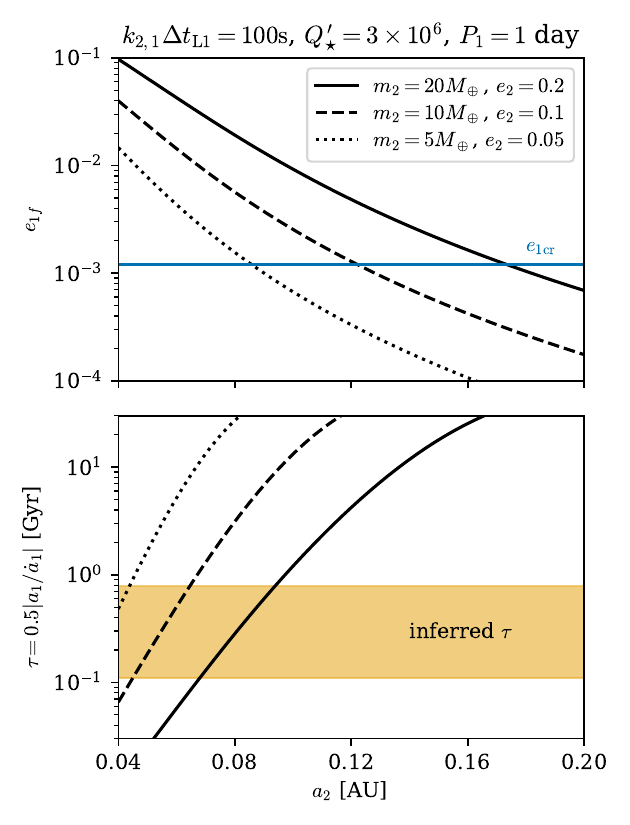}
    \caption{{\it Upper:} Forced eccentricity of a USP on a 1-day orbit 
    due to external companions of various masses $m_{2}$ and eccentricities $e_{2}$, 
    as a function of the companion's semi-major axis $a_{2}$. 
    Dissipation in the planet dominates its tidal evolution when $e_{\rm 1f}$ 
    exceeds the critical value $e_{\rm 1cr}$, marked with a blue horizontal line.
    {\it Lower:} The same for the estimated inspiral lifetime $\tau = 0.5 |a_{1}/\dot{a}_{1}|$ from a 1-day orbit. 
    The orange shaded region marks the range of derived USP lifetimes $\tau$ in Table \ref{tab:observations_to_Sig0_tau}. 
    We set $M_{\star} = \MSol$, $R_{\star} = \RSol$, $m_{1} = \ME$, $R_{1} = \RE$, $k_{2,1} \Delta t_{\rm L1} = 100 \, {\rm s}$, and $Q'_{\star} = 3 \times 10^{6}$ in all calculations.}
    \label{fig:inspiral_ef_adot}
\end{figure}

The top panel of Figure \ref{fig:inspiral_ef_adot} shows how $e_{\rm 1f}$  
depends on $a_{2}$ for an Earth-like USP around a solar-mass star with various companion masses and eccentricities. 
The bottom panel compares the estimated $\tau$ values (equation \ref{eq:tau_lifetime_estimate}) 
with the inferred $\tau$ values (Table \ref{tab:observations_to_Sig0_tau}). 
We consider companions with masses of $5$, $10$, and $20 \ME$; eccentricities of $0.05$, $0.1$, and $0.2$; 
and semi-major axes $a_{2}$ between $0.04$ and $0.2 \AU$. 
These span the normal range of masses and eccentricities among ``Kepler multis'' \citep[e.g.,][]{Xie+2016, VanEylen+2019}.
For $e_{\rm 1f} \ll e_{\rm 1cr}$, the USP's lifetime is set by stellar tides and exceeds $10 \Gyr$. 
Perturbations from companions of moderate mass ($\gtrsim 5 \ME$) and eccentricity ($\gtrsim 0.05$) and located at $a_{2} \lesssim 0.1 \AU$
are required to reproduce the inferred $\tau$ values. 

Unlike in Section \ref{s:Model:CompareUSPForm}, these results do not generalize to the other dynamical formation scenarios. 
Neither high-e migration \citep{PDW2019} nor obliquity-driven migration \citep{MS2020} 
provides a means for exciting the USP's eccentricity or obliquity at $P \leq 1$\,day. 
Driving a USP into its star on a timescale $\lesssim 1 \Gyr$ in either scenario 
would require stellar tidal friction with an effective $Q'_{\star} \ll 10^{6}$, 
which is unfeasible for low-mass planets orbiting MS stars \citep[e.g.,][]{Barker2020, MaFuller2021}. 
Only low-e migration can self-consistently account for both the $\Sigma_{0}$ and $\tau$ values we infer. 

\section{Discussion} \label{s:Discuss}

We find that USP engulfment is a natural consequence of the low-e migration scenario. 
A connection between USPs and engulfed rocky planets in Sun-like stars therefore seems plausible. 
We now discuss various implications of that connection for observations of USPs and their host stars, 
including identifying a few straightforward tests. 
We also address a few caveats.

\subsection{Occurrence of USPs, companions, and pollution} \label{s:Discuss:Implications}

Rocky USPs require secular forcing from outer companions to facilitate engulfment within $1 \Gyr$ of arrival. 
In the low-e migration scenario, the innermost companion 
must have $m_{2} \gtrsim 5 \ME$ and $a_{2} \approx 0.04$--$0.1 \AU$ ($P_{2} \approx 3$--$12$ days; see \citealt{PL2019}). 
A polluted star would be expected to have a planet meeting these criteria. 
That planet's geometric transit probability would be \citep{BoruckiSummers1984}
\begin{equation}
    \frac{R_{\star}}{a} \approx 0.1 \left( \frac{R_{\star}}{\RSol} \right) \left( \frac{a}{0.05 \AU} \right)^{-1}.
\end{equation}
Thus, we predict that in a suitable sample of Sun-like stars with chemical anomalies, 
5--10\% have compact systems with a transiting innermost planet having the aforementioned properties. 
Indeed, a similar fraction of known USPs have transiting companions \citep{Weiss+2018}. 
A corollary is that evidence of rocky planet engulfment in Sun-like stars 
should be \emph{anti}-correlated with the occurrence of USPs around those stars in the present day, 
because each planetary system forms one USP at most. 
It may be possible to test these predictions by cross-referencing the samples of stars examined by 
\citetalias{Spina+2021}, \citetalias{Behmard+2023b}, and \citetalias{Liu+2024} against an exoplanet database. 
This is beyond the scope of this Letter.

\subsection{USP host demographics} \label{s:Discuss:USPHostDemo}

\subsubsection{Ages} \label{s:Discuss:USPHostDemo:Ages}

It has previously been argued that USPs cannot have tidal inspiral timescales much shorter than $\sim 10 \Gyr$ \citep{HS2020}. 
This followed from the finding that the Galactic velocity dispersion of a sample of USP host stars 
-- and hence their relative ages -- is consistent with that of field stars. 
However, that argument implicitly assumes that USPs reach their observed orbits early in their host stars' lives. 
As we have discussed, the observed USP population may occupy a steady state, 
migrating inward gradually and decaying rapidly at sub-day periods due to tidal dissipation. 
Using our parameterized population model, we find the average USP host age is $\approx 3$--$6 \Gyr$, 
even if any given USP is destroyed by tidal inspiral on a timescale $\lesssim 1 \Gyr$. 
The old ages of USP host stars may reflect migration throughout MS lifetimes rather than lack of tidal evolution. 
Planet-engulfing stars should likewise have ages consistent with, or slightly older than, field stars. 

\subsubsection{Spectral types} \label{s:Discuss:USPHostDemo:SpType}

The USP occurrence rate varies with stellar spectral type:  $1.1\%$ of M dwarfs, $0.83\%$ of K dwarfs, $0.51\%$ of G dwarfs, and $0.15\%$ of F dwarfs 
have planets with orbital period $P < 1$ day and radius $R > 0.84 \RE$ \citep{SanchisOjeda+2014}. 
How does this trend fit into our scenario? 

The observed USP occurrence rate $\yuavg$ in our model is $\sim y_{p0} \tau / T$ (equation \ref{eq:define_yuavg_yzavg:yuavg_formula}), 
where $T$ is the maximum stellar age. 
However, $y_{p0}$ and $T$ should only be treated as constants when considering a population of stars with similar effective temperatures. 
To some extent, the decline in USP occurrence with increasing stellar mass must reflect 
the fact that small, short-period exoplanets are less abundant around earlier-type stars \citep[e.g.][]{Fressin+2013, Dressing2015} -- 
i.e., $y_{p0}$ should be expected to decrease with increasing stellar mass. 
Because the inner edges of protoplanetary disks lie at greater distances from hotter stars, 
there may be fewer planets poised for inward migration \citep{LC2017}. 

Our calculations in Section \ref{s:Model} suggest that the factor $\tau$ 
introduces an additional mass dependency $\propto t_{a1} / e_{\rm 1f}^{2} \propto M_{\star}^{4/3}$, 
(equations \ref{eq:adot_div_a_tides}, \ref{eq:e1forced}, and \ref{eq:define_epsGR}). 
To match the the trend of USP occurrence rates with spectral type in a low-e migration scenario, 
$y_{p0} \tau/T$ would need to decrease with increasing stellar mass overall. 

Further light could be shed on this issue by examining how the frequency of engulfed planets ($\sim \yzavg$) depends on stellar properties. 
Both \citet{Spina+2021} and \citet{Yong+2023} found the frequency of chemical anomalies in co-natal pairs
increases with the metal-enriched star's effective temperature. 
However, differences in stellar properties along the main sequence, 
such as the settling and mixing rates for heavy elements and the mass of the outer convection zone, 
complicate this finding (see Section \ref{s:Discuss:Caveats:settling}). 
We leave a more thorough examination of this question for future work. 

\subsection{Caveats} \label{s:Discuss:Caveats}

\subsubsection{Settling of heavy elements} \label{s:Discuss:Caveats:settling}

Planet engulfment signatures may disappear over time 
due to the settling of heavy elements out of a star's convective envelope. 
One-dimensional calculations find engulfment signatures last less than $\sim 1 \Gyr$ in most cases
\citep{Sevilla+2022, Behmard+2023b, Behmard+2023a}. 
One could incorporate this crudely into our model by adding a term $-y_{z}(t) / \tau_{\rm s}$ to equation (\ref{eq:diffeqs_yu_yz:yz}), 
where $\tau_{\rm s}$ is a predetermined settling timescale. 
However, one-dimensional treatments of atomic diffusion and thermohaline mixing in stars 
may overestimate settling rates and thus underestimate the duration of an engulfment signature 
(see \citealt{Liu+2024} and references therein). 

\subsubsection{Alternative routes to rocky planet engulfment} \label{s:Discuss:Caveats:Alternatives}

As mentioned above, violent dynamical evolution can inject planets into their stars. 
In Appendix \ref{app:violent}, we estimate that $\sim 2$--$3\%$ of FGK dwarfs could be polluted by exoplanets of any kind as a result of several violent channels in tandem; 
this is near the lower end of the observed pollution frequency range, close to the \citetalias{Behmard+2023b} value. 
However, the inferred composition and mass of the pollutants suggest that the engulfed bodies were rocky super-Earths specifically, 
rather than e.g.\ Jupiter- or Neptune-like worlds with greater masses or different chemical compositions. 
We find only $\sim 1\%$ of stars can be polluted through violent destruction of super-Earths, despite their ubiquity as exoplanets. 
Thus, if we allowed for violent destruction channels to contribute to the observed $\yzavg$ alongside USP engulfment, 
it would not greatly alter our inferred $\Sigma_{0}$ and $\tau$ values or the conclusions we draw from them. 

\subsubsection{What about hot Jupiters?} \label{s:Discuss:Caveats:HJs}

Hot Jupiters are believed to be destroyed by tidal inspiral within their hosts' MS lifetimes. 
Since these have a similar occurrence rate to USPs around Sun-like stars ($y_{\rm HJ} \approx 0.01$; e.g., \citealt{Mayor2011, Wright+2012}), 
one can ask how much their destruction contributes to chemical anomalies in this population. 

We examine whether HJ engulfment alone could account for the observed planet engulfment frequency using the model of Section \ref{s:Model}. 
We assume that HJs have an occurrence rate $\yuavg = y_{\rm HJ} = 0.01$ 
and that they migrate inward over stellar lifetimes 
from a long-period population with $y_{p0} = 0.1$ \citep[e.g.,][]{Wittenmyer+2020, Fulton+2021}. 
We find solutions for the \citetalias{Behmard+2023b} and \citetalias{Liu+2024} values for $\yzavg$  
with $\Sigma_{0} \approx \{ 0.11 , 0.22 \} \perGyrstar$ and $\tau \approx \{ 1.7 , 2.7 \} \Gyr$. 
A solution for \citetalias{Spina+2021}'s value is impossible because $\yzavg$ cannot exceed $y_{p0}$ (equation \ref{eq:define_yuavg_yzavg:yzavg_formula}). 
These values are plausible based on models of high-e migration for HJs \citep{DawsonJohnson2018}
and the expected dissipation in their host stars \citep{Hansen2012, Penev+2018, Barker2020}. 

However, there is reason to doubt a HJ contribution to pollution of Sun-like stars. 
Again, an engulfed HJ may not produce a similar chemical signature to a rocky planet: 
the masses and bulk metallicities of HJs vary widely \citep[e.g.,][]{Thorngren+2016}, 
and the hydrogen and helium content of the envelope may dilute extra metals in a stellar convection zone. 
Additionally, tidal disruption of a circularized HJ may not lead directly to engulfment. 
Theoretical studies predict stable mass transfer \citep{ValsecchiRasio2014, JiaSpruit2017}, 
potentially producing a super-Earth remnant comprising the original core and a residual envelope. 
This remnant is accreted many Gyr after the initiation of mass transfer \citep{ValsecchiRasio2014, Jackson+2016}. 
Thus, the signatures of engulfed HJ cores may appear mainly among older MS stars or subgiants. 

\section{Conclusion} \label{s:Conclusion}

We investigate a potential link between the tidal evolution of USPs 
and the evidence of rocky planet engulfment among Sun-like stars. 
Our main conclusions are: 
\begin{enumerate}
    \item[(i)] Rocky planet engulfment within the host's MS lifetime is a natural consequence of the low-e migration USP formation channel. 
    We suggest that the USP population among $\sim 1$--$10 \Gyr$-old Sun-like stars 
    reflects a quasi-steady state between inward migration via eccentric planetary tides 
    and subsequent tidal destruction. 
    
    \item[(ii)] We reproduce the USP occurrence rate and the frequency of engulfed planets 
    if the average USP formation rate among ``Kepler multis'' is $\Sigma_{0} \sim 0.01$--$0.2 \perGyrstar$ 
    and the average USP lifetime under tidal decay is $\tau \sim 0.1$--$0.8 \Gyr$. 

    \item[(iii)] USP lifetimes are mainly determined by tidal dissipation in the planet rather than the star. 
    This is because a USP's external companions can continually excite its eccentricity during its inspiral. 

    \item[(iv)] The host stars of observed USPs should have ages and kinematics consistent with Galactic field stars as a whole. 
    Rarely, if ever, should they have evidence of previous planet engulfment. 

    \item[(v)] Most polluted FGK dwarfs should host compact multi-planet systems. 
    The innermost planet should have a mass $\gtrsim 5 \ME$ and semi-major axis $\lesssim 0.1 \AU$, 
    transiting in $\sim 5$--$10 \%$ of cases. 
    
\end{enumerate}

Future work can gain further insights on USP formation and evolution by moving beyond treating $\Sigma$ and $\tau$ as constants, 
modeling the period distribution of USPs in a quasi-steady-state between formation and destruction, and 
revisiting the frequency of engulfment events as a function of stellar type.

\begin{acknowledgments}
We thank Jiaru Li, Sarah Millholland, and Fred Rasio for helpful discussions.
C.E.O.\ acknowledges support from a CIERA Postdoctoral Fellowship and National Science Foundation grant AST-2107796. 

\software{IPython/Jupyter \citep{PerezGranger2007_IPython, Kluyver+2016_Jupyter}, 
Matplotlib \citep{Hunter2007_matplotlib}, 
NumPy \citep{Harris+2020_NumPy}, 
SciPy \citep{Virtanen2020_scipy}, 
SymPy \citep{Meurer2017_SymPy}}
\end{acknowledgments}

\appendix

\section{Planet engulfment from violent dynamical evolution} \label{app:violent}

As discussed in the Introduction, pollution of Sun-like stars could be attributed to extrasolar planetary systems 
having experienced violent dynamical evolution that injected planets into their stars. 
In this Appendix, we compile and synthesize previously published results in the literature to quantify the feasibility of this explanation. 

In each of the following sections, we estimate two key quantities for various dynamical processes: 
$f_{\rm inj}$, the probability of injecting a planet into a star via a given mechanism; 
and $y_{\rm occ}$, the occurrence probability of a configuration susceptible to that mechanism. 
Together, these provide a simple estimate of the probability of observing a star that has engulfed a planet through that dynamical channel, 
denoted $y_{z}$ as in the main text: 
\begin{equation}
    y_{z} = f_{\rm inj} y_{\rm occ}.
\end{equation}

\subsection{Planet--planet scattering}

Close encounters as a result of dynamical instabilities can have three outcomes in general for a given planet: 
collision with another planet, usually resulting in a merger; ejection from the system; or injection into the central star. 
The branching ratios of these outcomes depend on various parameters, including the Safronov number
\begin{equation}
    \Theta = \frac{m}{M_{*}} \frac{a}{R}, 
\end{equation} 
where $m$ and $R$ are the combined mass and radius of two planets encountering one another 
at a distance $a$ from the central star of mass $M_{*}$. 
Analytical results describing necessary conditions for planet--star collisions may be found in \citet{Rodet2024}.

\subsubsection{Unstable gas giants}

Most numerical studies of planet--planet scattering have focused on unstable gas giants, 
mainly in two- or three-planet systems with $\Theta \sim 1$.
These generally find a small brancing ratio for injections (Table \ref{tab:scattering_results}) versus ejections or mergers. 
An illustrative value is $f_{\rm inj} = 0.05$. 
To estimate $y_{\rm occ}$ for systems with long-period giant planets (``cold Jupiters'' or CJs), 
we adopt the observed CJ occurrence rate as $y_{\rm CJ} = 0.1$--$0.2$ \citep[e.g.,][]{Wittenmyer+2020, Fulton+2021} 
and the fraction of CJs with additional companions as $y_{\rm comp} = 0.5$ \citep{Bryan2016}. 
Hence
\begin{equation}
    y_{z} = f_{\rm inj} y_{\rm CJ} y_{\rm comp} = 0.005 \left( \frac{f_{\rm inj}}{0.05} \right) \left( \frac{y_{\rm CJ}}{0.2} \right) \left( \frac{y_{\rm comp}}{0.5} \right).
\end{equation}
Although increasing the number of planets in a system makes dynamical instability more likely, 
the branching ratio for injections remains small (Table \ref{tab:scattering_results}).

\begin{table*}[]
    \centering
    \begin{tabular}{lccl}
        \hline\hline
        Source & No.\ Planets & Fraction & Notes \\ \hline 
        \citet{Ford2001} & 2 & $< 0.01$ & equal-mass planets \\
        \citet{Ford2008} & 2 & $0.03$ & planet-to-planet mass ratio $\beta = 0.5$ \\
        '' & '' & $0.12$ & $\beta = 0.3$ \\
        '' & '' & $0.16$ & $\beta = 0.2$ \\
        \citet{Petrovich2014} & 2 & $0.013$ & `2pl-fiducial', S $\cup$ (E+S) outcomes \\
        \citet{Li2021} & 2 & $< 0.02$ & tidal effects in close planet--planet encounters \\ \hline
        \citet{Chatterjee2008} & 3 & $0.02$ & \\ 
        \citet{Nagasawa2008} & 3 & $0.22$ & Set `N' \\
        \citet{Petrovich2014} & 3 & $0.029$ & `3pl-fiducial', 2S $\cup$ (C+S) $\cup$ (E+S) $\cup$ (C+E+S) \\
        '' & '' & $0.116$ & `3pl-a2', S $\cup$ 2S \\ 
        \citet{Anderson+2020} & 3 & $< 0.10$ & \\
        \hline
    \end{tabular}
    \caption{Fractions of unstable two- and three-giant-planet systems 
    that produce either direct star--planet collisions or planets on star-grazing orbits, 
    as obtained from selected numerical scattering experiments described in the literature. 
    Outcome definitions vary between studies, as do initial conditions.}
    \label{tab:scattering_results}
\end{table*}

\subsubsection{Unstable super-Earths}

The observed compact super-Earth systems are abundant ($y_{\rm occ} \approx 0.3$--$0.5$; e.g., \citealt{Fressin+2013, Zhu+2018})
and have likely experienced dynamical instabilities \citep{PuWu2015}. 
Since $\Theta \ll 1$ for these systems, close encounters lead largely to planet--planet collisions, 
with a negligible branching ratio for star--planet collisions. 
Hence we expect $y_{z} \ll 0.01$ for these systems even though $y_{\rm occ}$ is relatively large. 

\subsubsection{Super-Earths destabilized by migrating giants} 

\citet{Mustill2015} studied a scenario where a giant planet undergoing high-eccentricity migration 
disrupts a pre-existing compact super-Earth system. 
They write: ``The overwhelming outcome is that the [...] inner planets are destroyed, most commonly by collision with the star''; 
but they do not state the exact fraction. 
However, since $f_{\rm inj} \leq 1$, we obtain a illustrative constraint of $y_{z} \lesssim 0.01$ for this channel
based on the known $y_{\rm occ} \approx 0.01$ for ``hot'' and ``warm'' Jupiters \citep[e.g.,][]{Mayor2011, Wright+2012}. 
At least one polluted Sun-like star (XO-2N) hosts a hot Jupiter \citep{Ramirez+2015}, 
perhaps lending this channel some observational support. 

\subsection{Secular interactions}

\subsubsection{Lidov--Kozai mechanism}

Misaligned hierarchical triple systems experience secular eccentricity/inclination oscillations through the Lidov--Kozai mechanism. 
Of particular interest is the ``eccentric'' Lidov--Kozai (ELK) mechanism, 
which can excite eccentricities close to unity over a wide range of initial conditions \citep[e.g.,][]{Naoz2016}. 
This form of dynamical evolution can arise in several ways in exoplanet systems.

\subsubsection{Stellar companion}

The best-studied case of ELK in a planetary system involves 
a single planet with an initially circular orbit a few AU from its host star (near the ``ice line'') 
perturbed by a distant ($\sim 100$--$1000 \AU$) stellar companion. 
Using Monte Carlo methods (i.e.\ population synthesis) with reasonable ranges of planetary and companion properties and orbital architectures, 
various studies have found $f_{\rm inj} = 0.05$--$0.3$ for tidally disrupting a giant planet via ELK oscillations 
\citep[e.g.,][]{Naoz2012, Petrovich2015, Anderson+2016}. 
Analytical results illustrating how $f_{\rm inj}$ depends on various parameters are presented in \citet{Munoz2016}. 
We take $f_{\rm inj} = 0.2$ as a fiducial value.

The occurrence factor for this scenario is $y_{\rm occ} = y_{\rm ice} (1 - y_{\rm comp}) y_{\rm bin}$, 
where $y_{\rm ice}$ is the fraction of stars with a single planet beyond the ice line, 
$y_{\rm comp}$ is the fraction of those planets with a nearby substellar companion 
(which suppresses ELK oscillations), 
and $y_{\rm bin} \approx 0.07$ is the fraction of Sun-like stars that belong to a binary separated by $\sim 100$--$1000 \AU$ \citep{Raghavan2010}. 
With $f_{\rm inj}$ as above, $y_{\rm ice} = y_{\rm CJ} = 0.2$, and $y_{\rm comp} = 0.5$, we obtain $y_{z} \approx 0.0014$ for ELK injection of CJs. 
Including higher-multiplicity systems in $y_{\rm bin}$ increases the result to $0.002$. 

Neptune-sized planets are several times more common beyond the ice line than CJs \citep{Suzuki2016, Herman2019, Poleski2021}. 
All else being equal, setting $y_{\rm ice} = 0.5$ yields $y_{z} \approx 0.01$ 
for planetary disruption by ELK oscillations in stellar binaries. 
In principle, we could extend the estimate to even smaller planets, 
but their occurrence rates beyond the ice line are not well constrained. 

\subsubsection{Planetary companion}

Hierarchical, inclined two-planet systems are also prone to ELK oscillations \citep[e.g.,][]{Naoz+2011}. 
Forming systems with this architecture requires excitation of mutual inclinations by earlier planet--planet scattering or a close stellar flyby. 
We examine the latter case, the conditions for which have been systematically quantified: 
\citet{Rodet2021} calculated the fraction of two-planet systems in which the inner Jovian planet undergoes high-eccentricity migration 
via ELK oscillations induced by a stellar flyby in a young star cluster. 
We modify their equation (20) to obtain the fraction of stars that accrete a disrupted planet, 
instead of the fraction where the planet survives as a hot Jupiter: 
\begin{equation}
    y_{z} = 0.0025 \left( \frac{m_{\rm tot}}{2 \MSol} \right) \left( \frac{a_{2}}{50 \AU} \right) \left( \frac{y_{\rm CJ}}{0.2} \right) \left( \frac{y_{\rm comp}}{0.5} \right) \left( \frac{f_{\rm inj}}{0.2} \right) \left( \frac{n_{\rm cl}}{10^{3} \pc^{-3}} \right) \left( \frac{t_{\rm cl}}{20 \Myr} \right) \left( \frac{\sigma_{\rm cl}}{1 \, {\rm km \, s^{-1}}} \right)^{-1}.
\end{equation}
Here, $m_{\rm tot}$ is the total mass of the host and interloper stars; 
and $n_{\rm cl}$, $t_{\rm cl}$, and $\sigma_{\rm cl}$ are the number density, lifetime, and velocity dispersion of the cluster. 
The factor $\eta_{\rm initial}$ in \citeauthor{Rodet2021}\ corresponds to $y_{\rm occ}$. 
As before, we express this as $y_{\rm occ} = y_{\rm CJ} y_{\rm comp}$.  
It should be noted that the cluster's properties can vary widely. 

\subsubsection{Secular chaos}

Secular chaos occurs in systems of three or more planets due to the overlap of nonlinear secular resonances. 
Like the ELK effect, it readily causes the innermost planet's disruption with high eccentricity. 
The probability of this outcome is sensitive to the system's initial conditions, 
particularly the eccentricities, mutual inclinations, and orbital spacing. 
For definiteness, we adopt an injection probability $f_{\rm inj} \approx 0.1$ for systems experiencing secular chaos, 
as obtained by \citet{Teyssandier2019} 
for systems of three CJs with moderate eccentricities and mutual inclinations. 

A fraction $y_{\rm comp} = 0.5$ of all CJs have at least one outer companion of comparable mass \citep{Bryan2016}. 
If we assume (arbitrarily) the probability of two CJs having an additional companion also equals $y_{\rm comp}$, 
then $y_{\rm occ} = y_{\rm comp}^{2} y_{\rm CJ}$. 
Thus
\begin{equation}
    y_{z} \approx 0.0025 \left( \frac{f_{\rm inj}}{0.1} \right) \left( \frac{y_{\rm CJ}}{0.2} \right) \left( \frac{y_{\rm comp}}{0.5} \right)^{2}.
\end{equation}

Another case of interest is a system with two CJs and a single inner super-Earth, 
which can arise when the evolution of outer CJs triggers collisions in a compact inner system 
\citep[e.g.,][]{Huang2017, PuLai2021}. 
The frequency of engulfed planets in this case can be factored as 
\begin{equation}
    y_{z} = f_{\rm inj} y_{\rm CJ} y_{\rm comp} f_{\rm SE}^{(\rm CJ)} f_{1}^{\rm (SE)}.
\end{equation} 
The first three factors are as before. 
The quantity $f_{\rm SE}^{(\rm CJ)} \approx 0.9$ is the occurrence probability of super-Earths given the presence of CJs \citep{ZhuWu2018, Bryan2019}, 
and $f_{1}^{\rm (SE)} \lesssim 0.2$ is the fraction of super-Earth systems that contain a single planet \citep{Zhu+2018}. 
The final factor is important because compact multi-planet systems can effectively self-stabilize against perturbations from CJs \citep{Pu2018, Denham2019}. 
Thus
\begin{equation}
    y_{z} \approx 0.0018 \left( \frac{f_{\rm inj}}{0.1} \right) \left( \frac{y_{\rm CJ}}{0.2} \right) \left( \frac{y_{\rm comp}}{0.5} \right) \left( \frac{f_{\rm SE}^{\rm (CJ)}}{0.9} \right) \left( \frac{f_{1}^{\rm (SE)}}{0.2} \right).
\end{equation}

\subsection{Synthesis}

If all the channels above operated independently, the aggregated $y_{z}$ from violent dynamical evolution 
involving exoplanets of any kind could be as large as $\sim 3\%$. 
The greatest potential contributions, each at $\sim 1\%$, are from 
(1) ELK oscillations in wide binary systems involving Neptune-sized or larger planets 
and (2) giant planets destabilizing super-Earths during high-eccentricity migration. 
If we consider only single stars that engulf super-Earths, the upper bound is $\sim 1 \%$. 
This could match the observational upper limit reported by \citet{Behmard+2023b}, 
but not the results of \citet{Spina+2021} or \citet{Liu+2024}. 
Therefore, violent dynamical evolution alone is unlikely to explain the observed frequency of planet engulfment events. 

\bibliography{polluted_suns_ms_s1}
\bibliographystyle{aasjournals}

\end{document}